\documentclass[%
reprint,
superscriptaddress,
amsmath,amssymb,
aps,
prb,
]{revtex4-1}

\usepackage{graphicx}
\usepackage{dcolumn}
\usepackage{bm}
\usepackage{mathtools}
\usepackage{bbold}
\usepackage{color}

\begin{document}

\preprint{}

\title{Disorder and Critical Current Variability in Josephson Junctions}

\author{Miguel Antonio Sulangi}
\email[]{msulangi@ufl.edu}
\affiliation{
Department of Physics, University of Florida, Gainesville, FL, USA 32611
}

\author{T.A. Weingartner}
\affiliation{Department of Electrical and Computer Engineering,
 University of Florida, Gainesville, FL, USA 32611
}
\author{N. Pokhrel}
\affiliation{Department of Electrical and Computer Engineering,
 University of Florida, Gainesville, FL, USA 32611
}
\author{E. Patrick}

\affiliation{Department of Electrical and Computer Engineering,
 University of Florida, Gainesville, FL, USA 32611
}
\author{M. Law}
\affiliation{Department of Electrical and Computer Engineering,
 University of Florida, Gainesville, FL, USA 32611
}

\author{P. J. Hirschfeld}
\affiliation{
Department of Physics, University of Florida, Gainesville, FL, USA 32611
}
\date{\today}

\begin{abstract}
We investigate theoretically the origins of  observed variations in the critical currents of Nb/Al-AlO$_x$/Nb Josephson junctions in terms of various types of disorder. We consider the following disorder sources: vacancies within the Al layer; thickness variations in the AlO$_x$ layer; and ``pinholes'' (\emph{i.e.}, point contacts) within the AlO$_x$ layer. The calculations are all performed by solving the microscopic Bogoliubov-de Gennes Hamiltonian self-consistently. It is found that a small concentration of vacancies within the Al layer is sufficient to suppress the critical current, while the presence of a small number of thick regions of the oxide layer induces a similar effect as well. The pinhole scenario is found to result in anomalous behavior that resembles neither that of a pure tunnel junction nor that of a superconductor-normal-superconductor (SNS) junction, but a regime that interpolates between these two limits. We comment on the degree to which each of the three scenarios describes the actual situation present in these junctions.
\end{abstract}

\maketitle

\section{\label{sec:level1}Introduction}

Recent progress in the fabrication of nanoscale superconducting devices has led to a resurgence of interest in  superconducting electronics.\cite{McCaughan2014}  Maximal control over the properties of  individual elements  is desirable, primarily since even tiny variations in the properties of the  components represent potential disruptions in the overall operation of these circuits.  In fact, the large number of Josephson junctions incorporated into a single circuit in recent devices has reached the point where small statistical variations in the critical currents of these junctions have to be accounted for judiciously as part of good superconducting circuit design.\cite{tolpygo2016superconductor}

 For example, it is generally found that the critical currents of Nb/Al-AlO$_x$/Nb Josephson junctions\cite{gurvitch1983high} as grown by the MIT Lincoln Laboratory are found to vary considerably from junction to junction\cite{tolpygo2014fabrication}, although there is some debate as to whether these are distributed in a Gaussian or a non-Gaussian fashion.\cite{holmes2017non} In general these junctions are not very well-characterized with respect to defect type and concentration, although it is clear from transmission emission microscopy (TEM) that the materials are quite disordered, both within the barrier region and within the superconducting leads. The degree of disorder present is  not  well understood, however, and how the junction-to-junction variation depends on the sample growth process remains a mystery. While it is plausible that this observed variation in the critical currents is due to disorder, precisely {which} kind of defect types causes the variation is an open question. One can imagine that vacancies present in the barrier layer could result in the variation; indeed, one byproduct of the growth processes used to synthesize these junctions is the preponderance of vacancies in the oxide layer. Another possibility is that the oxide layer itself has thickness variations which lead to a nonuniform critical current density across the junction cross section. Finally, one possibility  frequently proposed is that  ``pinholes,'' are areas where the insulating oxide layer is almost completely absent, could explain these variations.\cite{tolpygo2014fabrication}

Here we demonstrate, using simplified, illustrative models that nevertheless capture the important disorder-driven physics of real-world Josephson junctions, that variations in the critical current can be accounted for by the presence of particular types of disorder. We use a microscopic approach in which the Bogoliubov-de Gennes (BdG) Hamiltonian is diagonalized self-consistently with the local superconducting order,\cite{blonder1982transition,furusaki1994dc,asano2001numerical,nikolic2002equilibrium,andersen2005bound,andersen20060,covaci2006proximity,andersen2008josephson,black2008self,graser2010grain,wolf2012supercurrent}, and the Josephson current across the  barrier for a given phase difference is then computed from the eigenvalues and eigenvectors of the BdG Hamiltonian. We will show that if vacancies (here modeled as spatially distributed strong local potentials) are present within a normal-metal barrier, the critical current is suppressed by a substantial amount relative to the clean case. We further consider the case where thick oxide regions are present within an otherwise thin barrier layer and find that as the number of thick regions increases, the reduction of the critical current is proportional to the exponential of the \emph{average} thickness of the oxide layer. Finally, we consider the pinhole scenario, where the barrier is dominated by a number of thin regions where the barrier transparency is large, and find that it gives rise to phenomena that do not correspond to tunnel junction behavior: the current-phase relation in the absence of impurities is sawtooth-like ---like that of a superconductor-normal-superconductor (SNS) junction---and even a small number of these pinholes is enough to enhance the critical current far beyond the clean tunnel-barrier limit. 

We argue that of the scenarios we considered, the likeliest explanation for the critical current variations observed in the MIT experiments is vacancies and small fluctuations (of the order of one lattice spacing) in the thickness of the oxide layer, as these are physically reasonable sources of disorder that are in fact seen in these junctions.  Reasonable statistical distributions of this type of disorder are shown to lead to distributions of critical currents consistent with that realized in experiments. On the other hand, we argue that pinholes are unlikely to provide a reasonable explanation for these. However, we make the case that pinholes are a likely origin of device failures, owing to the fact that even a tiny concentration of pinholes leads to critical currents much larger than the pinhole-free cases, and thus result in the failure of the junctions to work to their intended specifications.

It should be stressed that the list of disorder types we have considered is by no means exhaustive. It is known that grain boundaries suppress the critical current for high-$T_c$ superconductors,\cite{graser2010grain} and may play a role here as well. Hydrogen also readily dissolves into niobium\cite{kim2012direct}, settling into the interstitials of niobium's lattice structure; this results in the embrittlement of niobium.\cite{grossbeck1977low} Hydrogen interstitials can also act as a weak scatterers, generating a screened Coulomb potential whose effect is felt by the electrons indirectly, in a manner similar to the way dopants within the insulating layers in the cuprate high-temperature superconductors affect the electrons in the metallic layers.\cite{abrahams2000angle,zhu2004,nunner2005dopant, nunner2005microwave, nunner2006fourier,he2006local, scalapino2006relating,sulangi2017revisiting, sulangi2018quasiparticle} However, treating these forms of disorder with any sort of fidelity to the real world requires far more microscopic detail than we can presently access with Bogoliubov-de Gennes numerics. We thus postpone a discussion of grain boundaries and hydrogen interstitials to future work.

\section{Side-by-side comparison of different types of disorder}

\begin{figure}[t]
	\centering
	\includegraphics[width=\columnwidth]{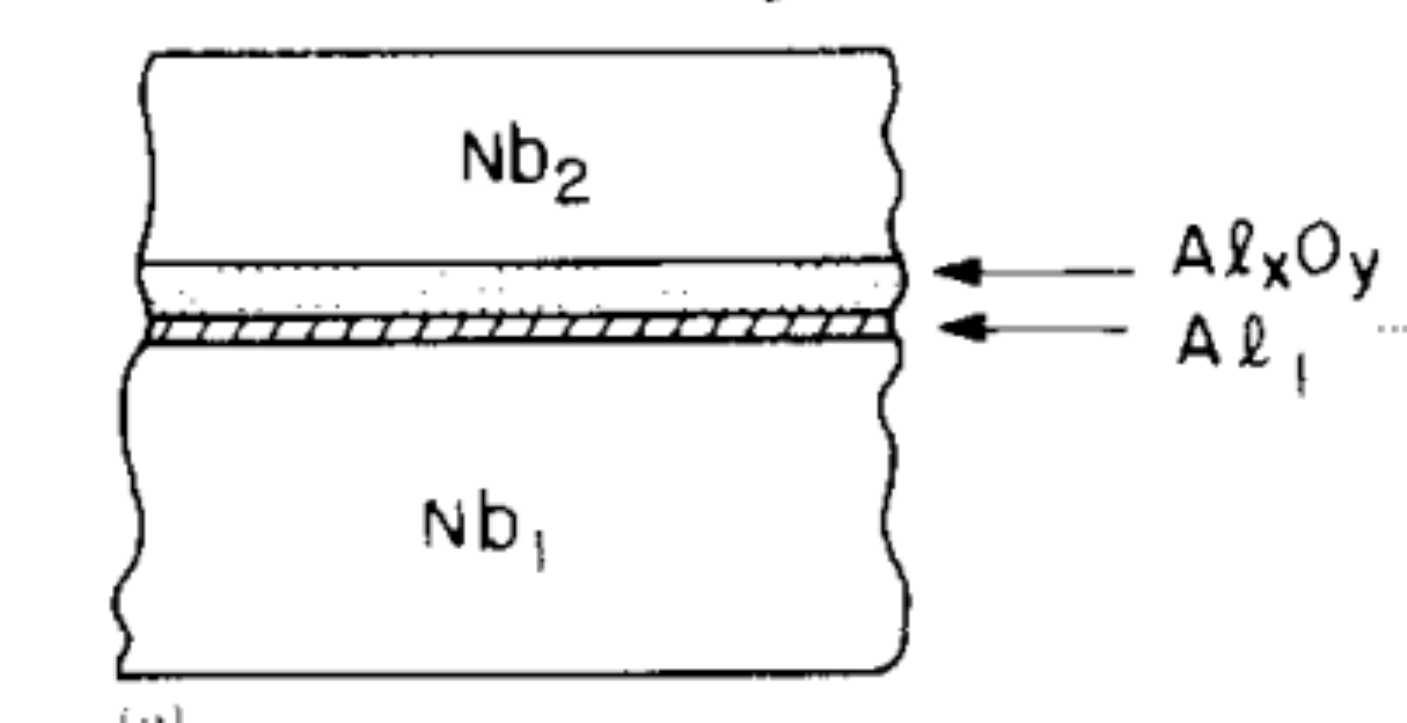}
	\caption{Schematic diagram of a Nb/Al-AlO$_x$/Nb junction, reproduced from Gurvitch \emph{et al.} (1983)\cite{gurvitch1983high}, with the permission of AIP Publishing. The Josephson junctions studied feature a base aluminum layer 8 nm thick, which oxidizes and forms an oxide layer (typically around 1 nm thick). The niobium leads are typically of the order of 100-200 nm in thickness.} 
	\label{fig:schematic}
\end{figure}

\begin{figure}[t]
	\centering
	\includegraphics[width=\columnwidth]{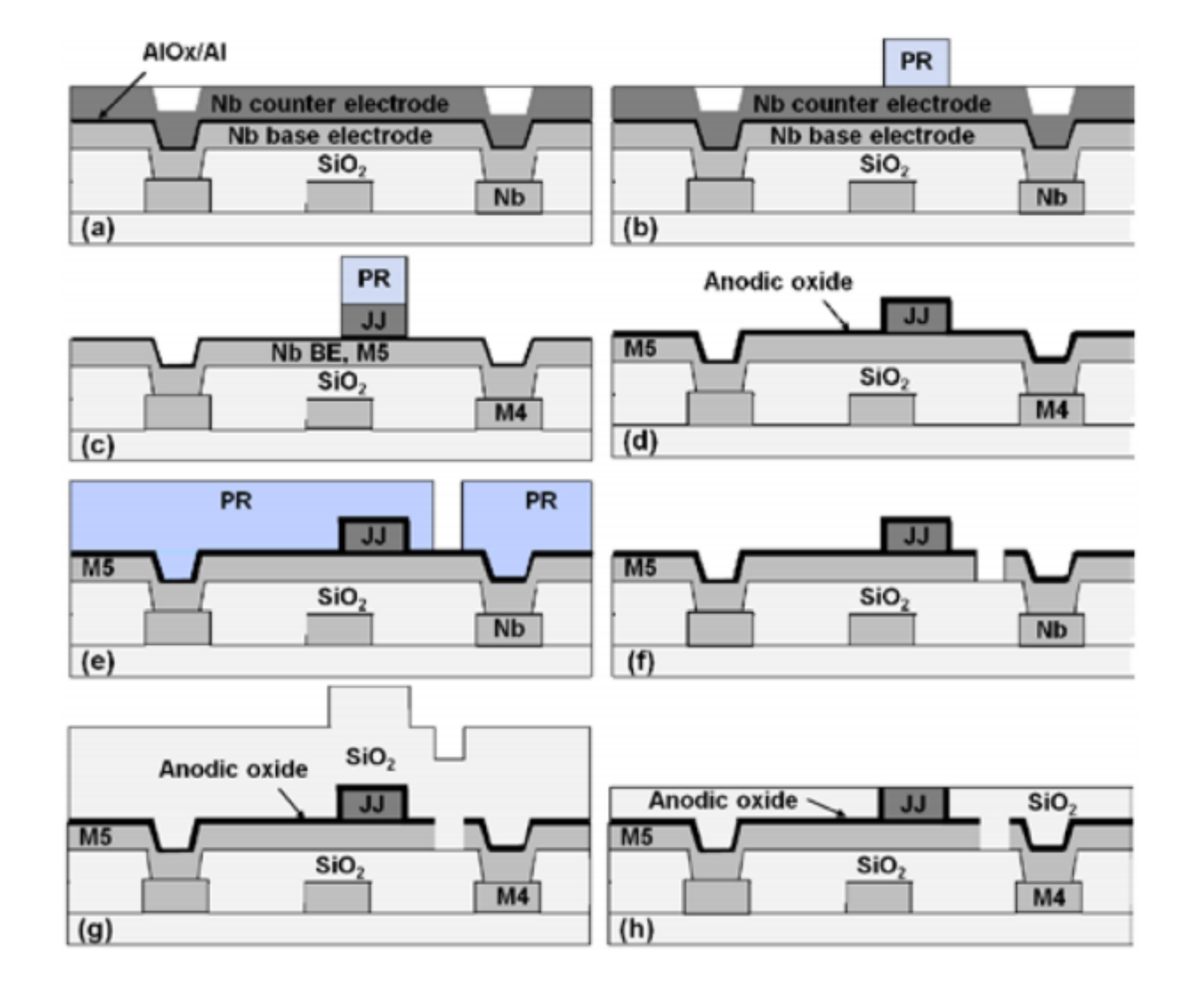}
	\caption{Fabrication process of a Nb/Al-AlO$_x$/Nb junction as employed by the MIT Lincoln Laboratory, reproduced from Tolpygo (2016)\cite{tolpygo2016superconductor}, with the permission of AIP Publishing. a) The Nb base electrode is deposited, then Al is deposited and subsequently oxidized to form an AlO$_x$ layer. A Nb counter-electrode is then placed above the Al-AlO$_x$ layer. b) A photoresist (PR) is placed above the counter-electrode. This sets the size of the Josephson junction. c) Photolithography is performed, resulting in the etching of the counter-electrode. The Al-AlO$_x$ layer is left exposed. d) Anodic oxidation is performed on all exposed surfaces of the Al-AlO$_x$ and the Nb counter-electrode to protect these from damage. e) Photoresists are placed to facilitate the etching of the Nb base electrode. f) The Nb base electrode is etched via photolithography. g) SiO$_2$ is deposited above the anodized surface. h) The SiO$_2$ layer is planarized via chemical mechanical polishing to the level of the top surface of the Nb counter-electrode of the Josephson junction. The full description of the process can be seen in Refs. \onlinecite{tolpygo2016superconductor} and  \onlinecite{tolpygo2014fabrication}.} 
	\label{fig:schematic2}
\end{figure}

To allow for an apples-to-apples comparison of the effect of different kinds of disorder, it is important first to fix the parameters of the system. The Nb/Al-AlO$_x$/Nb junction is modeled as a superconducting-normal metal-insulator-superconductor (SNIS) junction, and for simplicity the Al and AlO$_x$ layers are taken to be distinct from each other. (A schematic diagram of the grown junction is shown in Fig.~\ref{fig:schematic}, while the full fabrication process employed by the MIT Lincoln Laboratory is shown in Fig.~\ref{fig:schematic2}.) As Nb is a conventional superconductor, we take the order parameter to be $s$-wave.

For simplicity we assume that the full junction lives on a square lattice in two spatial dimensions, and hence the barrier is one-dimensional. The Hamiltonian describing the junction is given by the following standard expression:
\begin{equation}
\begin{split}
H =&-\sum_{\langle i, j \rangle} \sum_{\sigma} t_{ij}c_{i\sigma}^{\dagger}c_{j\sigma} + \sum_{\langle i, j \rangle}\big(\Delta_{ij}c_{i \uparrow}^{\dagger}c_{j \downarrow}^{\dagger} \\
&+ \Delta_{ij}^{\ast}c_{i \uparrow}c_{j \downarrow}\big).
\end{split}
\end{equation}
This can be recast into Bogoliubov-de Gennes form by applying a particle-hole transformation. The BdG eigenvalue problem can be written in the following more illuminating manner:
\begin{equation}
E_n
\left( \begin{array}{l}
u_n(\mathbf{r}_i)\\
v_n(\mathbf{r}_i)
\end{array} \right) =
\sum_j \\
\left( \begin{array}{ll}
-t_{ij} & \Delta_{ij}\\
\Delta^*_{ij} &  t_{ij}
\end{array} \right)
\left( \begin{array}{l}
u_n(\mathbf{r}_j)\\
v_n(\mathbf{r}_j)
\end{array} \right).
\end{equation}
Here $u_n$ and $v_n$ are the particle and hole amplitudes. Meanwhile the order parameter $\Delta_{ij}$ is obtained self-consistently from the eigenvalues and eigenvectors of $H$ using
\begin{equation}
\begin{split}
\Delta_{ij} =&\frac{V}{2} \delta_{ij}\sum_n \big[u_n(\mathbf{r}_i) v^*_n(\mathbf{r}_j)f(-E_n) \\
&-  v^*_n(\mathbf{r}_i) u_n(\mathbf{r}_j)f(E_n)\big].
\end{split}
\end{equation}
Note here that only the diagonal elements of $\Delta_{ij}$ are nonzero and that all its off-diagonal elements vanish, in keeping with the requirement that only $s$-wave superconductivity is present. Also note that within the normal metal and insulating barrier regions, the $s$-wave pairing interaction $V$ vanishes. To calculate the Josephson current, a phase gradient is applied across the system. The phase of the order parameter is held fixed at the left and right ends of the system with phases given by 0 and $\phi$, respectively, where $\phi$ is the phase difference across the junction; the phase in the rest of the system is allowed to fluctuate freely in the course of the self-consistent solution of the problem.

The current flowing between any two points $\mathbf{r}_i$ and $\mathbf{r}_j$ is computed using the following expression:
\begin{equation}
\begin{split}
j(\mathbf{r}_i, \mathbf{r}_j) =&-2i t_{ij} \sum_n \big[u_n(\mathbf{r}_i) u^*_n(\mathbf{r}_j)f(E_n) \\
&+  v_n(\mathbf{r}_i) v^*_n(\mathbf{r}_j)f(-E_n) - \text{h.c.}\big].
\end{split}
\end{equation}

The system consists of $30 \times 30$ lattice sites, with periodic boundary conditions along the $y$-direction and open boundary conditions along the $x$-direction. The $s$-wave pairing interaction $V = 2$ is chosen to ensure that the superconducting coherence length is parametrically larger than the thickness of the oxide barrier.  It is assumed that only nearest-neighbor hopping exists for the electrons (with $t = 1$), that the hopping amplitude is the same in both superconducting and normal regions, that both S and N are at half-filling, and that the temperature is in the limit $T \to 0$.

The Al  and AlO$_x$ layers are taken to be four lattice spacings and one lattice spacing thick, respectively. The thinness of the barrier (AlO$_x$) layer relative to the normal-metal (Al) region is chosen to mimic the small average thicknesses of the oxide layers seen in this particular class of Josephson junctions---for the junctions grown in particular by the MIT Lincoln Laboratory, the base aluminum layer from which the oxide is formed is around 8 nm thick\cite{tolpygo2016superconductor}, while the thickness of the oxide layer is generally of the order of 1 nm\cite{tolpygo2003tunneling, zeng2015direct}.  None of the phenomena described here in our simulations will turn out to be particularly sensitive to the thickness of the normal-metal layer. The oxide layer itself is modeled as a potential barrier of height $U = 4t$, where $t$ is the nearest-neighbor hopping amplitude. 

\begin{figure}[t]
	\centering
	\includegraphics[width=\columnwidth]{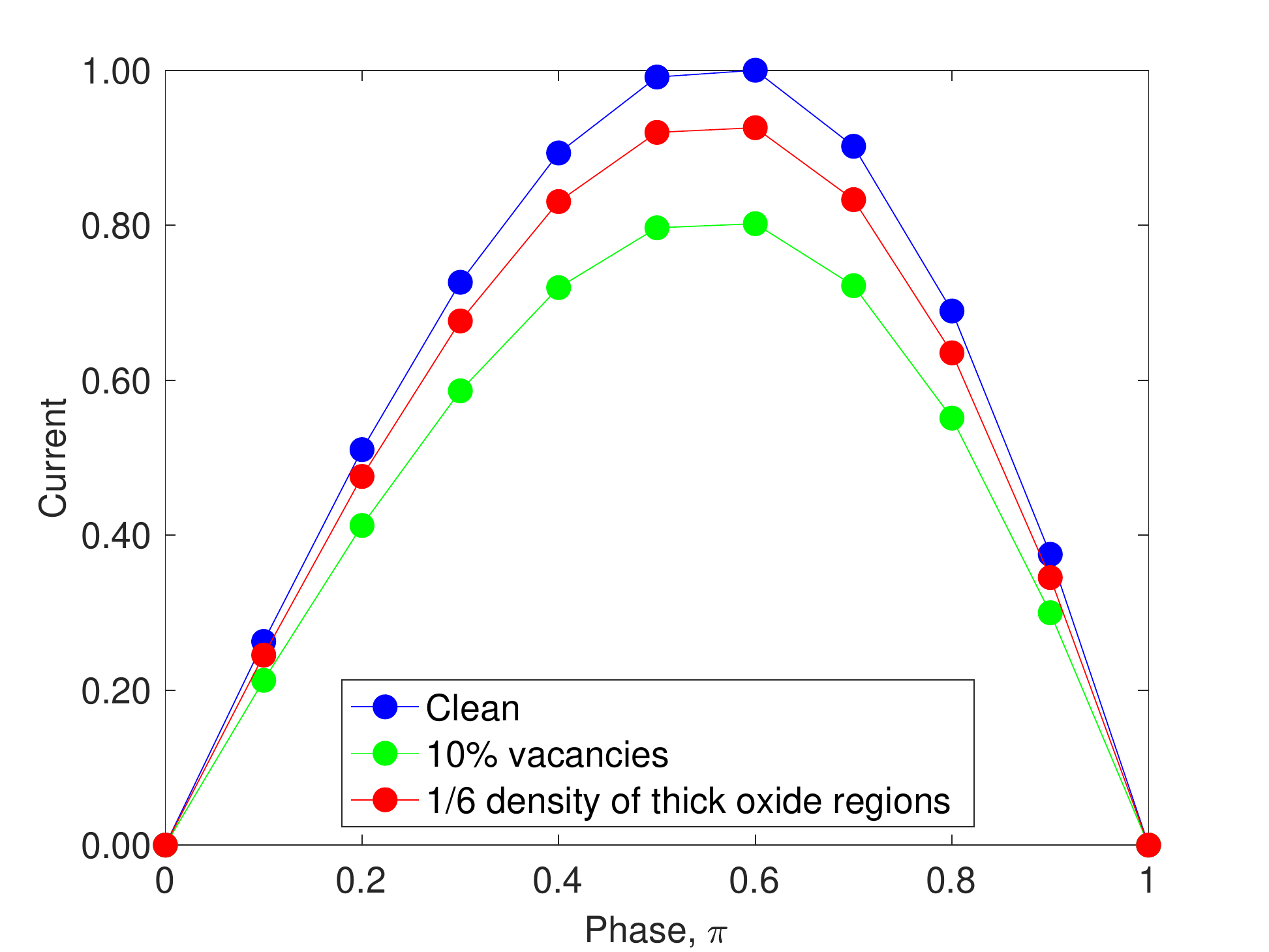}
	\caption{Current-phase relation of a SNIS junction with 10\% concentration of vacancies within the normal metal and 1/6 density of thick regions within the oxide layer, with the clean limit as a reference point. The currents shown are normalized relative to the critical current of the clean junction.} 
	\label{fig:CPR}
\end{figure}

\begin{figure}[t]
	\centering
	\includegraphics[width=\columnwidth]{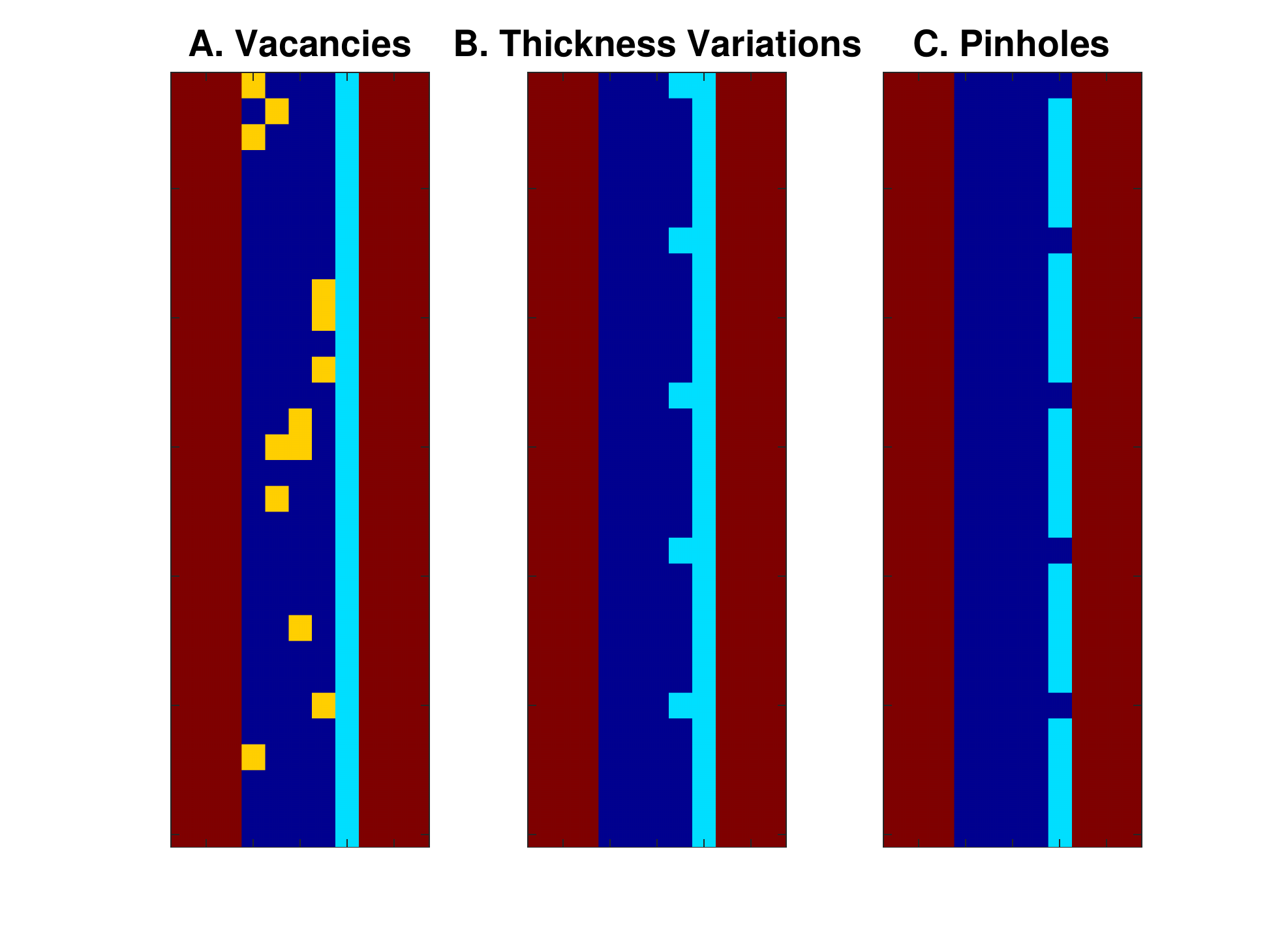}
	\caption{Disorder scenarios studied in this paper. Shown here are close-ups of the barrier region. In these plots, the following color scheme is used: blue = normal metal (\emph{i.e.}, the Al region), cyan = potential barrier (\emph{i.e.}, the AlO$_x$ layer), dark red = superconductor (\emph{i.e.}, Nb), and orange = vacancies. A.) An ensemble of randomly distributed strong impurities with approximate concentration of 10\%. B.) A periodic array of thick oxide regions (defined here as having a thickness of two lattice sites) distributed throughout the barrier layer. C.) A periodic array of pinholes (here defined as sites where the potential barrier is zero) distributed throughout the barrier layer.} 
	\label{fig:configs}
\end{figure}

Figure~\ref{fig:CPR} shows one of the main results of this paper. The current-phase relation (CPR) for a disordered N layer in an SNIS junction is plotted here, along with the CPR of the clean system to provide a baseline to which the disordered results can be compared. The CPRs corresponding to two particular disorder types are plotted here. The first is a randomly distributed ensemble of extremely strong on-site potentials (at a concentration of 10\%) localized within the normal-metal portion of the barrier, which models the presence of vacancies within the Al layer. The second model features an oxide layer whose thickness varies periodically across the barrier. Here five segments of the oxide layer have thickness equal to two lattice spacings, with the rest of the layer having the usual thickness of one lattice spacing. (The length of the barrier is 30 lattice sites, so the density of these thick regions is 1/6. Because periodic boundary conditions are applied along the barrier, to ensure commensurability and periodicity, only integer factors of 30 can be used in the number of thick oxide regions present.) Schematic depictions of the disordered systems studied in this section can be seen in Fig.~\ref{fig:configs}.

For both the clean and disordered cases, the CPR has a sinusoidal form, with a peak corresponding to the critical current near $\phi = \pi/2$. This is to be expected from a tunnel junction in the low-temperature limit.\cite{ishii1970josephson, golubov2004current} One can see that the main effect of disorder is to lower the critical current. For the 10\% concentration of vacancies, the critical current is approximately 20\% lower than that of the clean system. Meanwhile, for the periodically-varying oxide barrier with the aforementioned density of thick regions, the current is suppressed by around 7\% relative to the clean limit. In the succeeding section we will discuss in greater detail how the critical current depends on the defect concentration and the density of thick oxide regions.

\begin{figure}[t]
	\centering
	\includegraphics[width=\columnwidth]{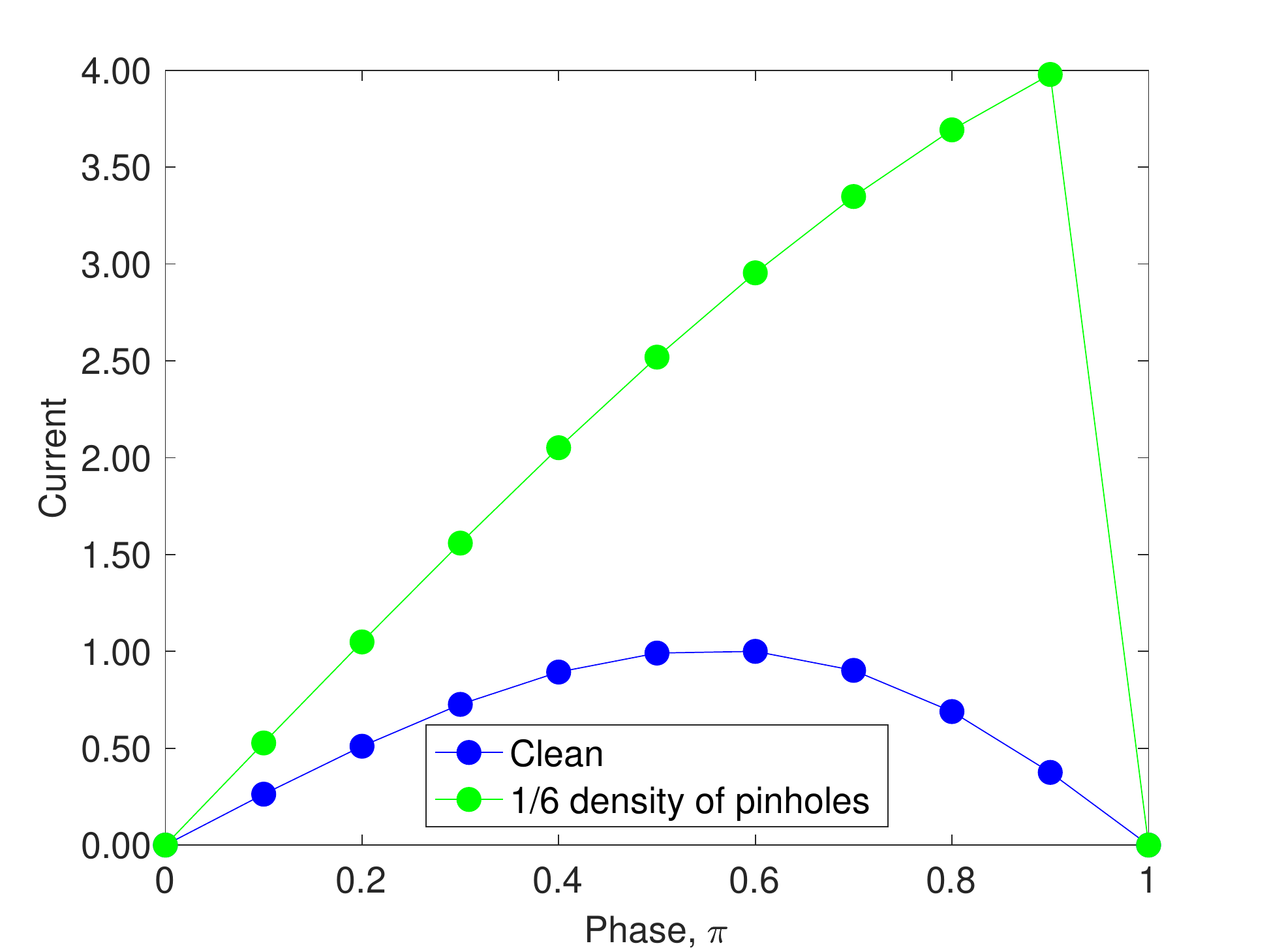}
	\caption{Current-phase relation of a SNIS junction with 1/6 density of pinholes within the oxide layer, with the clean limit as a reference point. The currents shown are normalized relative to the critical current of the clean junction.} 
	\label{fig:CPR_pinholes}
\end{figure}

The final case of pinholes is fascinating, not least because it turns out to behave completely differently from the other two forms of disorder we had considered. Figure~\ref{fig:CPR_pinholes} shows the CPR for a junction with five pinholes spread evenly across the barrier, which would correspond to a pinhole density of 1/6. One can see that the current for a given phase difference is markedly \emph{enhanced} relative to the clean tunnel-junction case. Even more strikingly, the CPR is completely altered from the sinusoidal form of the clean case. It now has a form resembling the sawtooth function. This is not surprising---the CPR of a clean SNS junction at $T \to 0$ is of a sawtooth form.\cite{golubov2004current} The alteration of the CPR reflects the changing nature of this junction, becoming less tunnel-junction-like and more SNS-like. In a later section we will discuss how increasing the number of pinholes interpolates between tunnel-barrier behavior and SNS-type behavior.

\section{Vacancies}

\begin{figure}[t]
	\centering
	\includegraphics[width=\columnwidth]{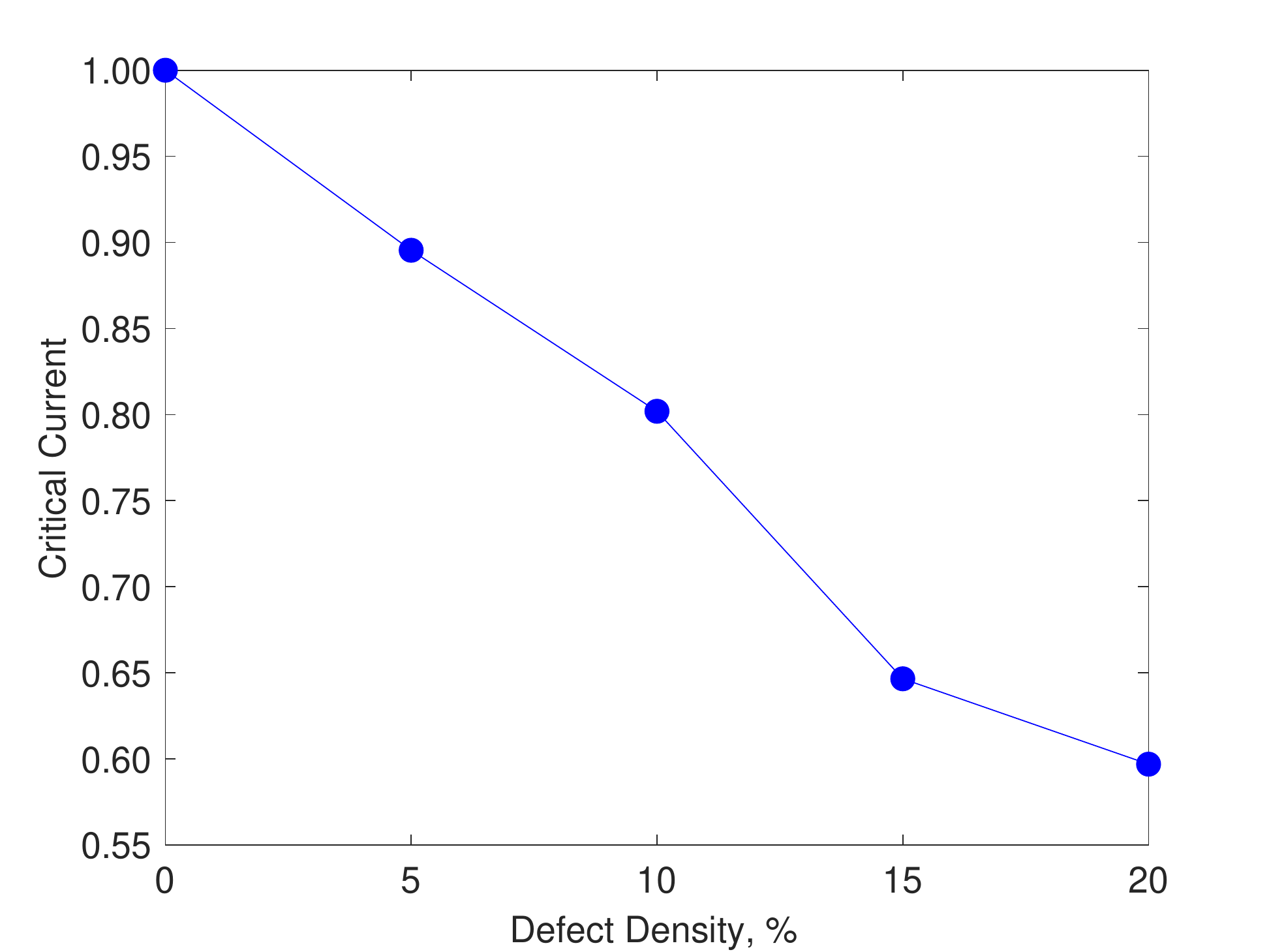}
	\caption{Critical current versus concentration of vacancies localized within the normal-metal portion of the barrier. A plot of one particular configuration of this disorder type (at 10\% concentration) can be seen in Fig.~\ref{fig:configs}A. The currents shown are normalized relative to the critical current of the clean junction (\emph{i.e.}, 0\% concentration).} 
	\label{fig:jj_vacancies}
\end{figure}

\begin{figure}[t]
	\centering
	\includegraphics[width=\columnwidth]{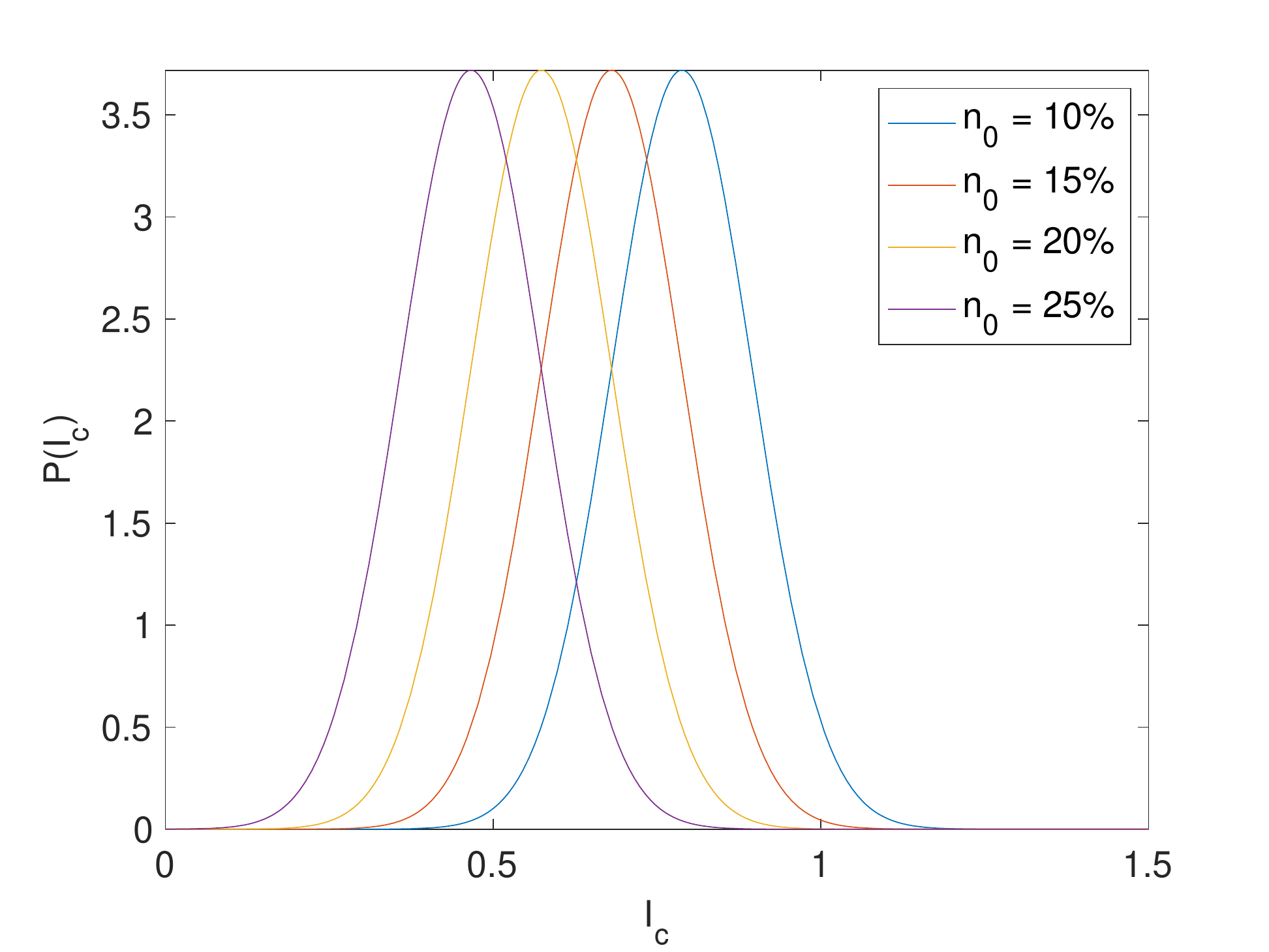}
	\caption{Distributions of critical currents for varying mean vacancy concentration $n_0$, assuming that the underlying impurity concentrations are normally distributed with standard deviation $\sigma = 5\%$.} 
	\label{fig:jj_vacancies_distributions}
\end{figure}

That the critical current is suppressed in the presence of strong on-site impurities within the Al layer should come as no surprise. The local density of states is strongly suppressed at sites which host such impurities, and as a result the effective amplitude for hopping from one site to an impurity site is decreased considerably. Thus the presence of strong impurities places a severe constriction on the available transport channels through which the Josephson current can pass.

The effect of impurity concentration on the critical current can be seen in Fig.~\ref{fig:jj_vacancies}, which shows that the critical current decreases monotonically as the impurity concentration is increased. There is a linear dependence of the critical current on impurity concentration at low concentrations. This linearity becomes weaker as the impurity concentration increases and impurity states begin to interfere. It is important to note that our results show that the suppression of the critical current does not follow a naive ``area-law'' expectation one might expect---namely that the critical current is proportional to the reduced effective area of the junction in the presence of vacancies, under which, say, a 10\% concentration of vacancies should give rise to a 10\% decrease in the critical current. The suppression in the presence of vacancies is in fact stronger than this effective area-law expectation, and is likely due to the real-space interference effects that are inevitably present in a randomly disordered system such as what we study.

It is important to note the large range over which the critical current varies (at 20\% concentration of vacancies, the critical current is down by almost 40\% relative to the clean limit). If we are to take this form of disorder as an explanation for the observed variations in the critical current, we have to ask if the sample fabrication process results in a highly variable number of defects present within the junctions. There is at present no good answer to this question. Given the very large number of samples that have been synthesized, the degree of junction-to-junction variation of the degree of disorder present remains largely unknown. We hypothesize that if the growth processes used yield a broad distribution of vacancy concentrations among all the samples synthesized, then this defect-centered explanation is very plausible. However, if it turns out that the processes result in a rather narrowly spread distribution of impurity densities, then any defect-based explanation would have to be reliant on specific junction-to-junction disorder configuration variations  \emph{at one concentration level}.\cite{beenakker1991universal}

This appears to be unlikely, however. One may ask if the critical current of a SNIS junction with a fixed concentration of vacancies is sensitive to the particular configuration of impurities present. We have checked this for a junction with a 10\% concentration of vacancies, performing simulations for nine different randomly generated configurations, and find that the standard deviation of the critical current of this ensemble of configurations is approximately 6\% of the mean value. This fixed-concentration variability is narrow compared to the variability arising from a varying concentration of defects. What this appears to suggest is the junction-to-junction variability seen in experiment is much more due to differences in the \emph{degree} of disorder present in each sample, as opposed to a situation wherein all samples have the same amount of disorder but have different disorder configurations present. This is a reasonable conclusion, especially in light of the fact that it is not known with certainty how much control the growth process has over junction-to-junction levels of disorder.

We can in fact make an estimate of the distribution of the critical currents if we assume that the junction-to-junction vacancy concentrations are distributed in some known fashion. As noted before, it can be seen in Fig.~\ref{fig:jj_vacancies} that the critical current depends on the impurity concentration $n$ in roughly a linear manner. If we assume that $n$ is itself normally distributed with mean $n_0$ and standard deviation $\sigma$, then $I_c(n) \propto n$ implies that $I_c$ is also normally distributed. This is more clearly illustrated in Fig.~\ref{fig:jj_vacancies_distributions}. Here we plot $P(I_c)$, the probability distribution function of $I_c$, for different mean impurity concentrations $n_0$. The linear dependence of $I_c$ on $n$ means that, as $n_0$ is varied, the centers of the distributions of $I_c$ are merely shifted, without altering the overall width of the distribution. Note that since we are interested primarily in the {\it shape} of the $I_c$ distribution, we have ignored additional physical broadening of the distribution due to disorder.

\section{Barrier-Thickness Variations}

\begin{figure}[t]
	\centering
	\includegraphics[width=\columnwidth]{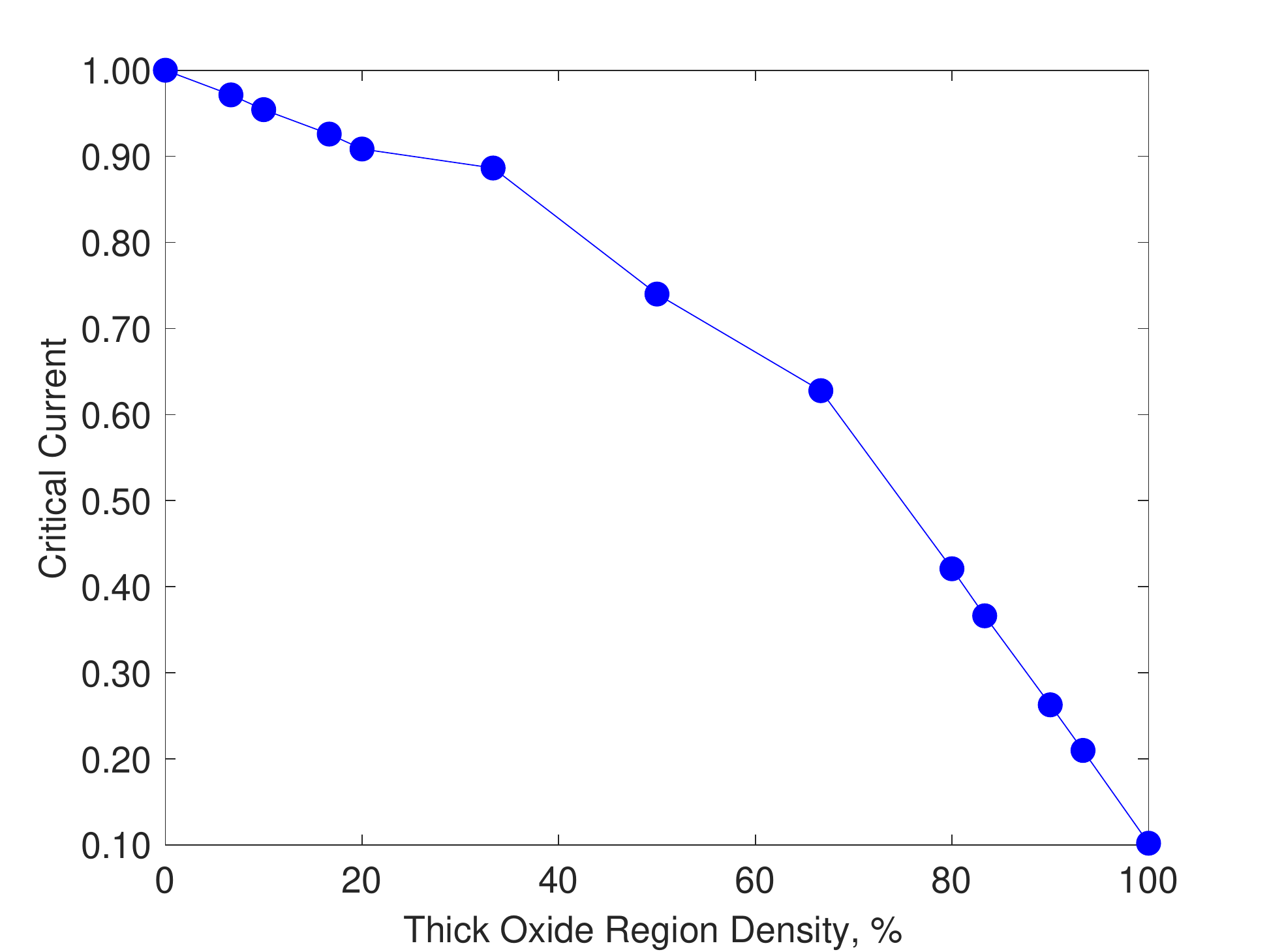}
	\caption{Critical current versus number of thick oxide barrier segments. The leftmost limit shows a uniformly thin barrier with thickness of one lattice spacing, while the rightmost limit shows a uniform barrier whose thickness is two lattice spacings. A plot of one particular configuration of this disorder type (17\%, corresponding to 5 thick segments) can be seen in Fig.~\ref{fig:configs}B. The currents shown are normalized relative to the critical current of the clean junction (with no thick regions present).} 
	\label{fig:jj_i_var}
\end{figure}

\begin{figure}[t]
	\centering
	\includegraphics[width=\columnwidth]{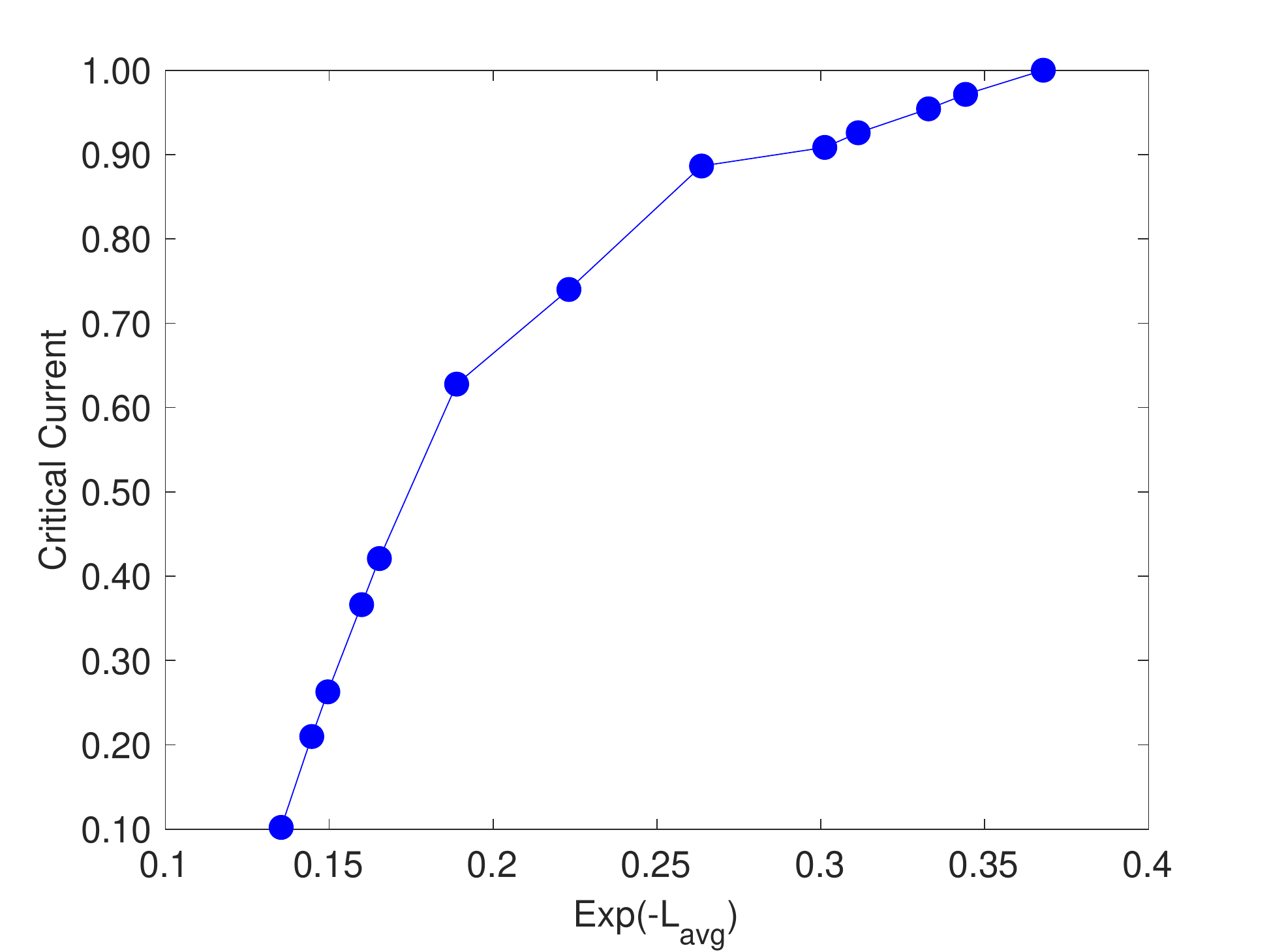}
	\caption{Critical current versus the exponential of minus the average oxide barrier thickness for the case of an oxide barrier with a spatially varying thickness. A plot of one particular configuration of this disorder type (17\%, corresponding to 5 thick segments) can be seen in Fig.~\ref{fig:configs}B.} 
	\label{fig:jj_exp_i_var}
\end{figure}

\begin{figure}[t]
	\centering
	\includegraphics[width=\columnwidth]{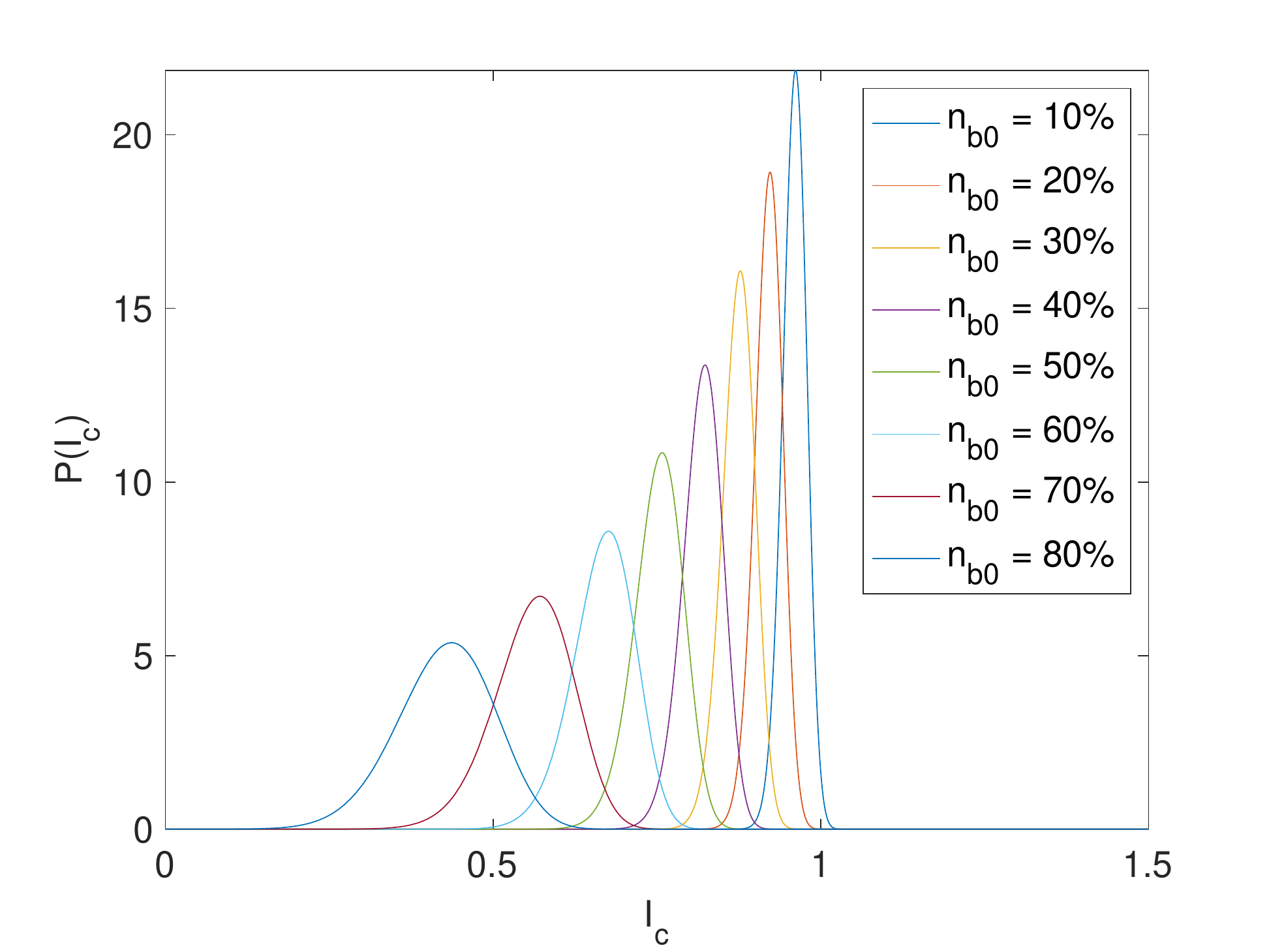}
	\caption{Distributions of critical currents for varying mean thick-oxide-segment concentration $n_0$, assuming that the underlying segement concentrations are normally distributed with standard deviation $\sigma = 5\%$.} 
	\label{fig:jj_i_var_distributions}
\end{figure}

The variation of the barrier thickness has frequently been invoked in the literature to explain the variability in the critical currents. From the numerics we have performed, the main mechanism by which the current is \emph{suppressed} in the presence of thick portions of the barrier region is the suppression of transport across the thicker regions of the oxide barrier. In Fig.~\ref{fig:jj_i_var} we show how the critical current depends on the number of thick regions present in the barrier. (As before, because of periodic boundary conditions along the barrier cross-section, we restrict the number of thick barrier sites to be integer factors of 30, the length of the cross section.) Notice that as the number of thick barrier sites is increased, the critical current decreases monotonically. The decrease is linear when the number of thick barrier sites is small, but becomes more pronounced downwards as their number becomes an appreciable fraction of the total number of sites across the cross section. 

It is interesting to see if, despite this atomic-scale variability in the barrier thickness, the critical current still has an exponential dependence on the barrier thickness. Figure~\ref{fig:jj_exp_i_var} shows how the critical current varies with the exponential of minus the \emph{average} barrier thickness. It can be seen that in the regime where the average thickness is small---hence a large exponential---the critical current depends linearly on $\exp(-L_{avg})$, where $L_{avg}$ is the average thickness of the oxide barrier. This functional dependence however becomes progressively less accurate with increasing average thickness, corresponding to a larger fraction of thick oxide barriers. Nevertheless the overall dependence of the critical current on \emph{average} barrier thickness---and not, for example, the minimum thickness---reflects the fact that the bulk superconducting coherence length (approximately 2-3 lattice sites for the parameters chosen in our simulations) is larger than the characteristic length scales of the oxide barrier region (which is one lattice size thick for the clean case), and hence the overall effect on bulk quantities such as the critical current is determined by averages over spatially fluctuating regions. 

We can make estimates for the distribution of the critical currents assuming knowledge of how the concentrations of thick-barrier sites are distributed on a junction-to-junction basis. Unlike the vacancies scenario, for oxide-barrier variations $I_c$ is a nonlinear function of the concentration of thick-barrier sites $n_b$, as one can see in Fig.~\ref{fig:jj_i_var}. For the purpose of computing $P(I_c)$ (the probability distribution function of $I_c$), we model the dependence of of $n_b$ on $I_c$ in the following way:
\begin{equation}
n_b(I_c) = Ae^{pI_c} + Be^{qI_c}.
\end{equation}
Here $A$, $B$, $p$, and $q$ are to be determined by a curve-fitting procedure. The precise functional form is not relevant---similar results can be obtained by a quadratic fit. Nevertheless with this functional form we can estimate $P(I_c)$, assuming that the thick-barrier site concentrations $n_b$ are normally distributed. Figure~\ref{fig:jj_i_var_distributions} shows the critical-current distributions as a function of the \emph{mean} thick-barrier site concentration $n_{b0}$. One can see that for small $n_{b0}$, the critical currents are narrowly distributed, whereas at large $n_{b0}$---the regime where a thick barrier region is present, punctuated by thin-barrier segments---the distributions are much broader. It can be noted that at intermediate and large $n_{b0}$ the critical-current distributions are skewed to the low critical current side, deviating from the symmetric Gaussian form present at small $n_{b0}$. These results show that the width of the critical-current distributions is determined to a large extent by whether the junction is in the thin-barrier regime with a small number of thick-barrier segments, or in the thick-barrier regime with a small concentration of thin-barrier sites. It has been argued that the Nb/Al-AlO$_x$/Nb junctions under study here indeed fall under the latter regime.\cite{holmes2019personal} It should be noted that the asymmetry seen here is opposite of what is seen experimentally in Nb/Al-AlO$_x$/Nb junctions, which have critical-current distributions that are skewed to the high $I_c$ side.\cite{holmes2017non} This suggests that other defects not considered here may be present and are responsible for this asymmetry. A possible explanation is a very small concentration of pinholes (discussed in greater detail in the following section) present in some junctions which cause the enhancement of the critical current, potentially leading to distributions that are skewed towards larger critical currents. Of course we also do not know if the concentrations of defects in different junctions are actually distributed in a Gaussian fashion; it is entirely possible that a different choice of distribution of $n_b$ could give rise to a  critical-current distribution skewed to the high $I_c$ side.  These results highlight the importance of obtaining independent measurements of the defect concentration in junctions produced by this process.  

\section{Pinholes}

\begin{figure}[t]
	\centering
	\includegraphics[width=\columnwidth]{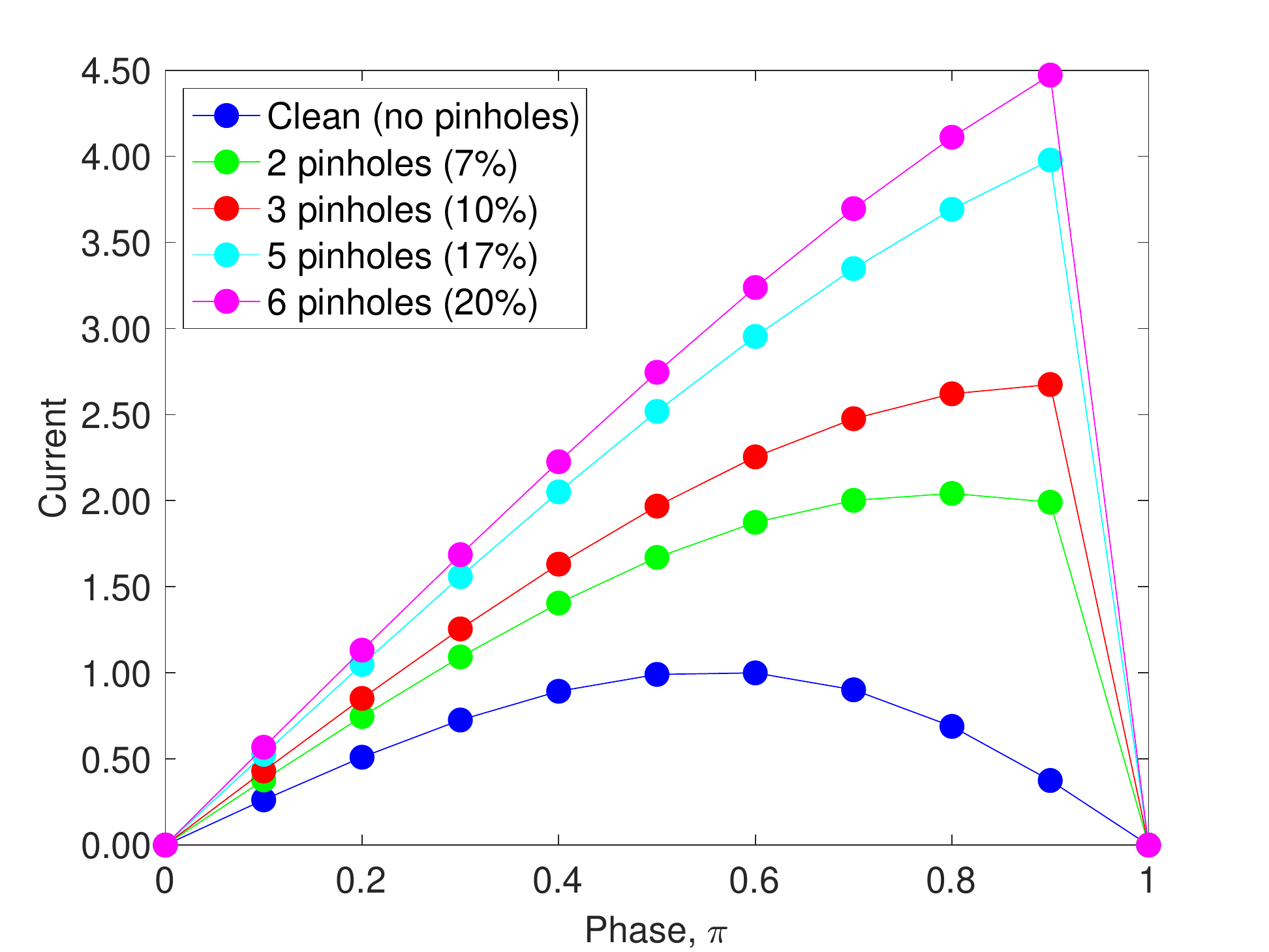}
	\caption{Current-phase relation of SNIS junctions with a varying number of pinholes present. The currents shown are normalized relative to the critical current of the clean junction.} 
	\label{fig:CPR_pinholes_full}
\end{figure}

\begin{figure}[t]
	\centering
	\includegraphics[width=\columnwidth]{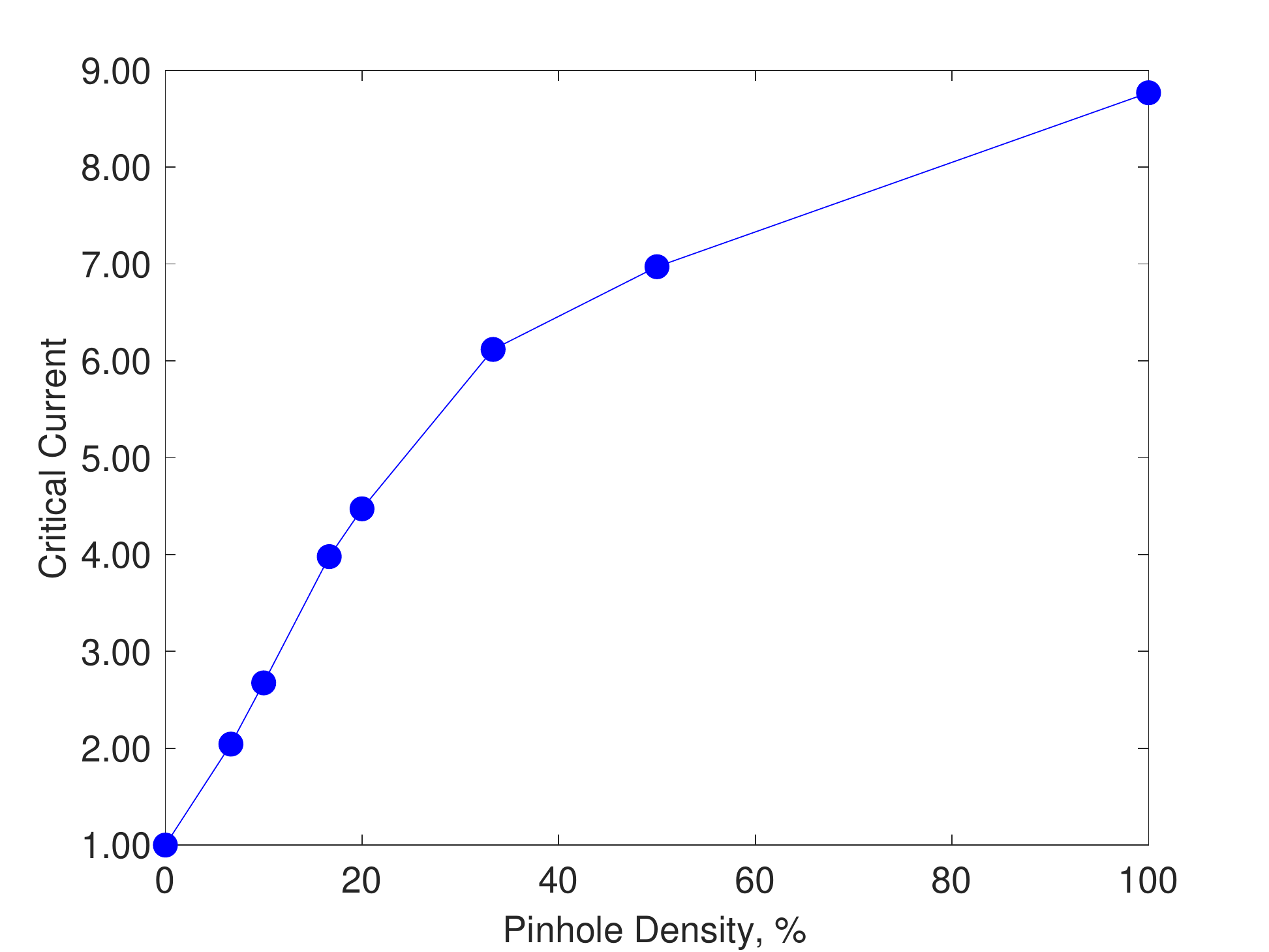}
	\caption{Critical current versus number of pinholes. The leftmost limit shows a uniformly thin barrier with thickness of one lattice spacing, while the rightmost limit shows an SNS junction with no insulating barrier. A plot of one particular configuration of this disorder type (with 5 pinholes) can be seen in Fig.~\ref{fig:configs}C. The currents shown are normalized relative to the critical current of the clean junction with no pinholes.} 
	\label{fig:jj_pinholes}
\end{figure}

\iffalse
\begin{figure}[t]
	\centering
	\includegraphics[width=\columnwidth]{"jj_exp_pinholes".pdf}
	\caption{Current at $\phi = \pi/2$ versus the exponential of minus the average oxide barrier thickness for the case of pinholes. A plot of one particular configuration of this disorder type (with 5 pinholes) can be seen in Fig.~\ref{fig:configs}C.} 
	\label{fig:jj_exp_pinholes}
\end{figure}
\fi

It is presently not well understood if the process employed by the MIT Lincoln Laboratory spontaneously generates pinholes in the fabricated junctions, and presently available information from TEM measurements suggests that these are not present in functioning junctions. Therefore much of this section is speculative and driven mainly by the academic question of what the effects of a nonzero concentration of pinholes are, as well as how the critical currents of junctions depend on the concentration of pinholes present in the barrier. We have shown earlier how a junction with only a 1/6 density of pinholes in the thin oxide layer already shows a sawtooth-like CPR, which is also seen in SNS junctions. How would such junctions behave as the number of pinholes is increased from zero? In Fig.~\ref{fig:CPR_pinholes_full} we show the CPRs for junctions with a varying number of pinholes present. One can see that even for a small (\emph{e.g.}, 2-3) number of pinholes, the CPR already displays a marked deviation from the sinusoidal form characteristic of the tunnel junction. The phase at which the supercurrent is maximum can be seen to shift from $\pi/2$ towards $\pi$ as the number of pinholes present is increased. The CPR becomes more and more linear with increasing pinhole density, reflecting the fact that as the barrier becomes more transparent its behaves more similarly to an SNS junction than a tunnel junction. This behavior is in fact strikingly similar to that theoretically predicted for a simplified one-dimensional model of a point contact with increasing barrier transparency.\cite{haberkorn1978theoretical} 

In hindsight these results are not very surprising. Intuitively, the supercurrent flowing through a tunnel junction with pinholes can be decomposed into two parts---one comes from supercurrent through the tunnel barrier, while the other arises from supercurrent flow through the pinholes. The former is sinusoidal, while the latter has a sawtooth form. In this picture, increasing the pinhole density should make the CPR more sawtooth-like. Our numerical results confirm this picture, and suggest that any possible real-space interference effects present in a junction with a pinholes spread out over the tunnel barrier do not appear to lead to significant deviations from this intuitive picture.

In Fig.~\ref{fig:jj_pinholes} we show the critical current as a function of pinhole number. Here we took the maximum value of the supercurrent as the phase is varied to be the critical current, with the phase difference at which this occurs shifting from $\phi = \pi/2$ to $\phi = \pi$ with increasing pinhole density.  One can see that the critical current is a monotonically increasing function of the number of pinholes, in agreement with expectations---the pinholes allow the unobstructed transport of supercurrent across the junction, and therefore should enhance the total amount of current. The dependence of the critical current on the number of pinholes is almost perfectly linear at small pinhole densities, with the nonlinear effects setting in around the $\sim 20 \%$ level. 

It is worth noting that the critical currents for junctions with even a small number of pinholes present are much larger than in the pinhole-free case---witness how a 10\% concentration of pinholes gives rise to a critical current \emph{twice} as large as in the clean case---with the critical current scaling approximately linearly with the pinhole concentration. Of all the forms of disorder studied in this paper, pinholes are the only type which gives rise to an \emph{enhancement} on the order of the clean-junction critical current itself, while the other two disorder types result in critical currents whose \emph{suppression} relative to the clean case is comparatively small. The variability of the critical current seen in experiment is small relative to the mean value of the critical current, and appears to suggest that vacancies or oxide-barrier variations are far likelier explanations for the aforementioned variation than pinholes.

\section{Discussion}

With the knowledge of the qualitative effects of various kinds of disorder on the critical current of SNIS Josephson junctions, one can make educated guesses as to the causes of the variability of the measured critical currents. One important factor that must be considered when attributing this effect to disorder is the variability of the junction-to-junction disorder levels. For instance, the presence of oxygen vacancies in the barrier regions  depends crucially on details of the growth process such as the exposure time. In general, there is no indication that the disorder in these junctions is ``controlled'' in the sense that \emph{fixed} concentrations of, for example, vacancies, dopants, or barrier fluctuations can be induced in a systematic way. Disorder arising here from the growth process appears to be intrinsic, but to our knowledge there is no \emph{quantitative} information, besides the critical-current variations, about the level of disorder present. Even estimates of junction properties such as the barrier thickness depend on averaging out the spatial variations that are intrinsic to these materials.\cite{tolpygo2003tunneling} Thus, in the absence of more atomic-level information about the spatial inhomogeneities present in these materials one cannot conclude much about how disorder-related properties such as vacancy concentrations or barrier-thickness variability are distributed across a large population of Josephson junctions.

Thus vacancies and barrier-thickness variations are excellent explanations for the critical-current variability in these junctions. The main appeal of these explanations is that these two forms of disorder are known to be present in these materials, and are intrinsic byproducts of the growth processes used for these junctions. It remains to be seen if the growth processes result in broadly distributed levels of disorder across various samples, as this would account for much of the observed junction-to-junction variability of the critical current to a greater degree than a narrowly-distributed disorder source would (for which the resulting junction-to-junction critical-current fluctuations  would be less distributed and far more reliant on rare-region effects present within one sample). 

Pinholes by themselves on the other hand appear to be unlikely explanations for the observed variability in the critical current. The reason is that the critical-current variability is much more uncontrolled and of a larger order of magnitude than seen in experiments. As mentioned earlier, the presence of even a small density of pinholes within the oxide barrier results in critical currents whose magnitude is far greater than in the clean-junction case. Therefore, when considered as an explanation for the critical-current distributions reported in the literature, pinholes would appear to result in a much wider distribution than seen in reality.  

However, it has to be noted that the reported distributions pertain only to \emph{functioning} devices. If the fabrication process results in some fraction of junctions having a finite concentration of pinholes within their respective oxide-layer barriers, then one can imagine a scenario in which junctions with pinholes let through much larger currents than the other components of the devices are capable of, owing to the large barrier transparency present, and therefore result in \emph{device failures}. It is not known what the yield is for the process employed in fabricating these junctions. From the point of view of optimizing the yield of the fabrication process, an interesting direction is to revisit the non-functioning devices and examine if these have pinholes present within the oxide barrier. If pinholes are indeed the culprit behind low yields, it would be worthwhile to modify appropriately the fabrication process in order to reduce the probability that device failures occur.  A direct measurement of the current-phase relationship of these junctions, e.g. with a SQUID, would be an independent test for the existence of pinholes.

We have also seen that the current-phase relation of the junction at low temperatures is an important indicator of the presence of pinholes, with high-transparency (\emph{i.e.}, pinhole-infested) junctions showing a sawtooth-like CPR, as opposed to a sinusoidal one. One possible future direction is to directly measure the current-phase relation on these junctions to see if these indeed conform to the expectations seen theoretically for SNIS junctions with a nonzero concentration of pinholes.\cite{il1998nonsinusoidal,il1999anomalous,il2000current,gotz2000supercurrent,il2001degenerate,il2001measurement,il2001radio,komissinski2002observation,grajcar2002supercurrent,grajcar2002superconducting} Understanding precisely what happens to these junctions by measuring the CPR, as well as revisiting TEM images of these junctions---especially junctions which result in device failures---should go a long way in clarifying their composition, as well as providing insight into improving their synthesis for future, large-scale applications.

\begin{acknowledgments}
We thank D. S. Holmes and A. Wynn for useful discussions. This work was supported by the IARPA SuperTools program.  PJH received partial support from NSF-DMR-1849751.
\end{acknowledgments}

\nocite{*}

\bibliography{report}

\end{document}